\documentclass[preprint]{revtex4}
\usepackage{graphicx}
\usepackage{bm}
\usepackage{dcolumn}
\usepackage{amsmath}
\usepackage{amsmath}
\usepackage{amssymb}
\usepackage{color}
\usepackage{verbatim}
\usepackage[toc,page]{appendix}

\usepackage{epsfig}
\begin{document}
\newcommand{\newc}{\newcommand}
\newc{\ra}{\rightarrow}
\newc{\lra}{\leftrightarrow}
\newc{\lsim}{\buildrel{<}\over{\sim}}
\newc{\gsim}{\buildrel{>}\over{\sim}}
\title{Updated reduced CMB data and constraints \\
on cosmological parameters }
\author{Rong-Gen Cai}
\email{cairg@itp.ac.cn}
\author{Zong-Kuan Guo}
\email{guozk@itp.ac.cn}
\author{Bo Tang}
\email{tangbo@itp.ac.cn}

\affiliation{
   State Key Laboratory of Theoretical Physics, Institute of
Theoretical Physics, Chinese Academy of Sciences, P.O. Box 2735,
Beijing 100190, China  }
\date{\today}

\begin{abstract}

We obtain the reduced CMB data $\{l_A, R, z_*\}$ from WMAP9, WMAP9+BKP, Planck+WP and Planck+WP+BKP for the $\Lambda$CDM
and $w$CDM models with or without spatial curvature. We then use these reduced CMB data in combination with low-redshift observations to put constraints on cosmological parameters.
We find that including BKP results in a higher value of the Hubble constant
especially when the equation of state of dark energy and curvature are allowed to vary.
For the $\Lambda$CDM model with curvature, the estimate of the Hubble constant with Planck+WP+Lensing
is inconsistent with the one derived from Planck+WP+BKP at about 1.2 $\sigma$ confidence level.
\end{abstract}

\maketitle
\section{Introduction}

Since the discovery of the cosmic acceleration expansion of the universe
based on the distance measurement of type Ia Supernovae (SNe Ia)~\cite{Riess:1998cb,Perlmutter:1998np},
its origin has become a
hot topic in modern cosmology and theoretical physics. The cause to the
observed cosmic acceleration is due to the so-called dark energy with negative
pressure in general relativity framework, or the modification to
 general relativity at cosmic scales.

To study properties of dark energy, one may combine some mature probes, such as SNe Ia,
the observational Hubble parameter (HUB), baryon acoustic oscillation (BAO)
and cosmic microwave background (CMB) anisotropy. The SNe Ia, HUB and
BAO probe the expansion of the universe at low and intermediate redshifts, while the CMB
measurements probe the distance at high-redshift (especially the distance to
the surface of last-scattering). The CMB data provide the strongest
constraints on cosmological parameters \cite{Komatsu:2010fb} and help
break the degeneracies among the dark energy and other cosmological parameters.

The reduced CMB data ${\{l_A, R, z_*\} }$ provide an efficient summary of CMB information,
of which $l_A$ is the angular scale of the sound horizon at recombination and determines the acoustic peak structure of the CMB angular power
spectra, $R$ is the scaled distance to the recombination and determines the amplitude of the acoustic peak and
$z_*$ is the redshift at the last scattering surface. Instead of the full CMB spectra, the reduced CMB data relating the distance to
the last scattering surface provide a fast and self-consistent approach for combining the CMB information with
complementary cosmological data to constrain late-time cosmological parameters.
The reduced CMB data were firstly derived in~\cite{Wang:2007mza} from the three-year WMAP data for the $\Lambda$CDM model (with and without spatial curvature) and they
found that dark energy density is consistent with a constant in cosmic time and
a flat universe is allowed by using the reduced CMB data together with the SNe Ia and BAO data.
Recently Wang and Wang~\cite{Wang:2013mha} obtained the reduced CMB data from WMAP9 and Planck+WP+Lensing for the
$\Omega_k$+$\Lambda$CDM model and found that the reduced CMB data
derived from Planck+WP+Lensing are much tighter than those from WMAP9, but when combined with
other low-redshift observational data, the reduced CMB data do not improve the constraints on the dark
energy too much compared to those from WMAP9. Besides, Shafer and Huterer~\cite{Shafer:2013pxa} derived
the reduced CMB data from WMAP9 and Planck+WP respectively for
the flat $w$CDM model, when combining the reduced CMB data, BAO and
3 samples of SNe Ia data (Union2.1~\cite{Union2.1}, SNLS3~\cite{Conley:2011ku} and PS1~\cite{Rest:2013mwz})
respectively, they found that there is a preference for
the equation of state of dark energy $w<-1$ for the constraints with Planck but not with WMAP9.
More recently,  the reduced CMB data have been obtained  based on the flat $w$CDM model by use of
the Planck 2015 temperature and low-$\it l$ polarization data, showing that the reduced CMB data are consistent with those
based on $\Lambda$CDM and  CPL  models~\cite{Ade:2015rim}.
 There are
many works that use the reduced CMB data together with complementary cosmological data to constrain late-time
cosmology~\cite{Su:2011ic,Cai:2013owa,Gao:2013pfa,Zhang:2013hza}.

 There are two advantages of the reduced CMB data.
First, the CMB shift parameters $l_A$ and $R$ together with the decoupling redshift $z_*$, extracted from the CMB angular power spectra,
allow one to quickly evaluate the likelihood of various dark energy models, without the need to run a Markov Chain Monte
Carlo exploration of the CMB likelihood
which usually includes a number of astrophysical parameters to describe unresolved foreground components
and other nuisance parameters.
Second, it provides an efficient and appropriate summary of CMB data as far as dark energy constraints are concerned.
Since $l_A$ determines the acoustic structure in CMB angular power spectra while $R$ determines the overall amplitude of the acoustic peaks,
they are nearly uncorrelated.
Both $R$ and $l_A$ can be used to further compress CMB information and combined with other measurements
in a friendly user manner to constrain dark energy models.

On the other hand, the authors of~\cite{Corasaniti:2007rf} analyzed the likelihood of the reduced CMB data
with WMAP3 data for the base $\Lambda$CDM model involving extra parameters, such as tensor modes and a running
spectral index. They found that adding curvature or slightly modifying the dark energy parameters does
not significantly change the values of $\{l_A, R\}$, which, however, change  large when   more
parameters like tensor modes or running of the scalar spectral index are involved.

The purposes of this work are to update constraints on the parameter combination \{$l_A, R, z_*$\}
using newly released CMB temperature and polarization data for several cosmological models
and to test their dependence on model assumptions.
These updated reduced CMB data provide a simple and efficient method for combining in a friendly
user manner the current CMB measurements with low-redshift data.
We first obtain the reduced CMB data from WMAP9 data~\cite{Hinshaw:2012aka}
and Planck data~\cite{Ade:2013zuv} together with WMAP polarization data (WP),
based on the $\Lambda$CDM model and $w$CDM model with a flat
or curved space curvature, respectively.
We also make a joint analysis of the data from
BICEP2/Keck Array~\cite{Ade:2015fwj} and Planck (BKP) to derive the reduced CMB data.
 Our goal is to see the differences among the
data used when combining the reduced CMB data with the low-redshift observational data to constrain different
cosmological models.

The paper is organized as follows. In section~\ref{sec:model} we present the
reduced CMB data obtained from WMAP9, Planck+WP, Planck+WP+BKP and WMAP9+BKP
based on different cosmological models, respectively.
In section~\ref{sec:results} we give the results of the combination of
reduced CMB data and other data sets to constrain the different cosmological models.
The conclusions are included in section~\ref{sec:conclusions}.


\section{Reduced CMB data \label{sec:model}}

The distance measurement is one of the most powerful methods to study the evolution history  of the universe.
In a Friedmann-Robertson-Walker universe, the comoving distance from an observer to redshift $z$ is given by
\begin{equation}
\begin{split}
&r(z)= H_0^{-1}  \left| \Omega_k \right|^{-1/2} {\rm sinn} [\left| \Omega_k \right|^{-1/2}\Gamma(z)],\\
&\Gamma(z)=\int_0^z \frac{dz'}{E(z')},~~E(z)=H(z)/H_0,
\label{eq:3}
\end{split}
\end{equation}
where $\Omega_k=-k/H_0^2$ ($k$ is the spatial curvature constant) and ${\rm sinn}(x)=\sin(x),x,\sinh(x)$ for $\Omega_k<0$, $\Omega_k=0$, and $\Omega_k>0$, respectively.
The Hubble parameter is given by the Friedmann equation
\begin{equation}
\begin{split}
H^2(z)=H^2_0[\Omega_{r0}(1+z)^4+&\Omega_{dm0}(1+z)^3+\Omega_{b0}(1+z)^3+\\
       &\Omega_{k}(1+z)^2+(1-\Omega_{m0}-\Omega_{r0}-\Omega_{k})],
\label{eq:1}
\end{split}
\end{equation}
for the $\Lambda$CDM model, where the redshift $z$ is defined by
$(1+z)=1/a$, and $\Omega_{r0}$, $\Omega_{dm0}$ and $\Omega_{b0}$ are
the present values of the fraction energy density  for radiation,
dark matter and baryon matter, respectively. The latter two are
often written as the total matter density
$\Omega_{m0}=\Omega_{b0}+\Omega_{dm0}$. The radiation density is the
sum of photons and relativistic neutrinos~\cite{Hinshaw:2012aka}:
\begin{equation}
\Omega_{r0}=\Omega^{(0)}_{\gamma}(1+0.2271N_{eff}),
\end{equation}
where $N_{eff}=3.046$ is the effective number of neutrino species  in the Standard Model of particle physics~\cite{Mangano:2005cc}, and $\Omega^{(0)}_{\gamma}=2.469\times10^{-5}h^{-2}$ for $T_{\rm CMB}=2.725K$
($h \equiv H_0/100$ km s$^{-1}$ Mpc$^{-1}$).
For the $w$CDM model, the Hubble parameter is given by
\begin{equation}
\begin{split}
H^2(z)=H^2_0[\Omega_{r0}(1+z)^4+&\Omega_{dm0}(1+z)^3+\Omega_{b0}(1+z)^3+\\
      &\Omega_{k}(1+z)^2+(1-\Omega_{m0}-\Omega_{r0}-\Omega_{k})F(z)],
\label{eq:2}
\end{split}
\end{equation}
where the evolving function $F(z)$, depending on the equation of state of dark energy, is given by
\begin{equation}
F(z)=(1+z)^{3+3w}.
\end{equation}

It is noticed that here we have not assumed a flat universe model. As for a flat universe,
the curvature terms disappear in equation~(\ref{eq:1}) and equation~(\ref{eq:2})~
since $\Omega_k=0$. There are four cases considered here, which are the $\Lambda$CDM model
and the $w$CDM model with a flat and curved space curvature, respectively.

\begin{table}[!tbh]
\begin{tabular}{l|l|l|l|l}
\hline Data  & flat $\Lambda$CDM  & $\Omega_k$+$\Lambda$CDM  & flat $w$CDM  & $\Omega_k$+$w$CDM  \tabularnewline \hline
\hline WMAP9      & this work & Ref.~\cite{Wang:2013mha} & Ref.~\cite{Shafer:2013pxa} & Ref.~\cite{Hinshaw:2012aka} \tabularnewline \hline
\hline PLANCK+WP  & this work & Ref.~\cite{Wang:2013mha} (+Lensing) & Ref.~\cite{Shafer:2013pxa} & this work \tabularnewline \hline
\hline PLANCK+WP+BKP & this work & this work & this work & this work  \tabularnewline \hline
\hline WMAP9+BKP & this work & this work & this work & this work  \tabularnewline \hline
\end{tabular}
\caption{\label{tab1} References for the reduced CMB data derived from different CMB data in different cosmological models.}
\end{table}

\begin{table}[!th]
\newcommand{\tabincell}[2]{\begin{tabular}{@{}#1@{}}#2\end{tabular}}
  \centering
\begin{tabular}{l|l|l|l|l}
\hline Data & $l_A\pm\sigma$    &   $R\pm\sigma$        & $z_*\pm\sigma$     & Correlation~Matrix \tabularnewline \hline
\hline WMAP9      & $301.95\pm 0.66$  &   $1.7257\pm 0.0165$  & $1088.96\pm 0.84$  &
  $\begin{bmatrix}
  1.0000&0.3859&0.4998\\
  0.3859&1.0000&0.8432\\
  0.4998&0.8432&1.0000
  \end{bmatrix}$ \tabularnewline \hline
\hline PLANCK+WP      & $301.66\pm 0.18$  &   $1.7500\pm 0.0089$  & $1090.33\pm 0.53$  &
  $\begin{bmatrix}
  1.0000&0.5126&0.4552\\
  0.5126&1.0000&0.8699\\
  0.4552&0.8699&1.0000
  \end{bmatrix}$ \tabularnewline \hline
\hline $\tabincell{c}{PLANCK+WP\\+BKP}$      & $301.61\pm 0.18$  &   $1.74974\pm 0.0087$  & $1090.04\pm 0.53$  &
  $\begin{bmatrix}
  1.0000&0.5526&0.4851\\
  0.5523&1.0000&0.8725\\
  0.4851&0.8725&1.0000
  \end{bmatrix}$ \tabularnewline \hline
\hline WMAP9+BKP      & $301.83\pm 0.66$  &   $1.7210\pm 0.0165$  & $1088.61\pm 0.85$  &
  $\begin{bmatrix}
 1.0000&0.4112&0.5217\\
  0.4112&1.0000&0.8550\\
  0.5217&0.8550&1.0000
  \end{bmatrix}$ \tabularnewline \hline
\end{tabular}
\caption{\label{tab2}The mean values, standard deviations of $\{l_A, R, z_*\}$ and the
         correlation matrix for the flat $\Lambda$CDM model.}
\end{table}

\begin{table}[!th]
\newcommand{\tabincell}[2]{\begin{tabular}{@{}#1@{}}#2\end{tabular}}
  \centering
\begin{tabular}{l|l|l|l|l}
\hline Data & $l_A\pm\sigma$    &   $R\pm\sigma$        & $z_*\pm\sigma$ or $w_b\pm\sigma$     & Correlation~Matrix \tabularnewline \hline
\hline WMAP9      & $302.02\pm 0.66$  &   $1.7327\pm 0.0164$  & $0.02260\pm 0.00053$  &
  $\begin{bmatrix}
  1.0000&0.3883&-0.6089\\
  0.3883&1.0000&-0.5239\\
  -0.6089&-0.5239&1.0000
  \end{bmatrix}$ \tabularnewline \hline
\hline $\tabincell{c}{PLANCK+WP\\+Lensing}$      & $301.57\pm 0.18$  &   $1.7407\pm 0.0094$  & $0.02228\pm 0.00030$  &
  $\begin{bmatrix}
  1.0000&0.5250&-0.4475\\
  0.5250&1.0000&-0.6925\\
  -0.4475&-0.6925&1.0000
  \end{bmatrix}$ \tabularnewline \hline
\hline $\tabincell{c}{PLANCK+WP\\+BKP}$      & $301.56\pm 0.19$  &   $1.7416\pm 0.0097$  & $1089.74\pm 0.59$  &
  $\begin{bmatrix}
 1.0000&0.5681&0.5279\\
  0.5681&1.0000&0.8946\\
  0.5279&0.8946&1.0000
  \end{bmatrix}$ \tabularnewline \hline
\hline WMAP9+BKP      & $301.94\pm 0.66$  &   $1.7251\pm 0.0167$  & $1088.88\pm 0.89$  &
  $\begin{bmatrix}
 1.0000&0.4112&0.5217\\
  0.4112&1.0000&0.8550\\
  0.5217&0.8550&1.0000
  \end{bmatrix}$ \tabularnewline \hline
\end{tabular}
\caption{\label{tab3}The mean values, standard deviations of $\{l_A, R, z_*\}$ and the
         correlation matrix for the $\Omega_k$+$\Lambda$CDM model.}
\end{table}

\begin{table}[!th]
\newcommand{\tabincell}[2]{\begin{tabular}{@{}#1@{}}#2\end{tabular}}
  \centering
\begin{tabular}{l|l|l|l|l}
\hline Data & $l_A\pm\sigma$    &   $R\pm\sigma$        & $z_*\pm\sigma$     & Correlation~Matrix \tabularnewline \hline
\hline WMAP9      & $301.98\pm 0.66$  &   $1.7302\pm 0.0169$  & $1089.09\pm 0.89$  &
  $\begin{bmatrix}
  1.0000&0.4077&0.5132\\
  0.4077&1.0000&0.8580\\
  0.5132&0.8580&1.0000
  \end{bmatrix}$ \tabularnewline \hline
\hline PLANCK+WP      & $301.65\pm 0.18$  &   $1.7499\pm 0.0088$  & $1090.41\pm 0.53$  &
  $\begin{bmatrix}
  1.0000&0.5262&0.4708\\
  0.5262&1.0000&0.8704\\
  0.4708&0.8704&1.0000
  \end{bmatrix}$ \tabularnewline \hline
\hline $\tabincell{c}{PLANCK+WP\\+BKP}$      & $301.65\pm 0.18$  &   $1.7495\pm 0.0087$  & $1090.31\pm 0.51$  &
  $\begin{bmatrix}
  1.0000&0.5379&0.4782\\
  0.5379&1.0000&0.8659\\
  0.4782&0.8659&1.0000
  \end{bmatrix}$ \tabularnewline \hline
\hline WMAP9+BKP      & $301.88\pm 0.66$  &   $1.7227\pm 0.0174$  & $1088.69\pm 0.92$  &
  $\begin{bmatrix}
  1.0000&0.4293&0.5267\\
  0.4293&1.0000&0.8722\\
  0.5267&0.8722&1.0000
  \end{bmatrix}$ \tabularnewline \hline
\end{tabular}
\caption{\label{tab4}The mean values, standard deviations of $\{l_A, R, z_*\}$ and the
         correlation matrix for the flat $w$CDM model.}
\end{table}

\begin{table}[!th]
\newcommand{\tabincell}[2]{\begin{tabular}{@{}#1@{}}#2\end{tabular}}
  \centering
\begin{tabular}{l|l|l|l|l}
\hline Data & $l_A\pm\sigma$    &   $R\pm\sigma$        & $z_*\pm\sigma$     & Correlation~Matrix \tabularnewline \hline
\hline WMAP9      & $302.40\pm 0.67$  &   $1.7246\pm 0.0183$  & $1090.88\pm 1.00$  &
  $\begin{bmatrix}
  1.0000&0.4262&0.5391\\
  0.4262&1.0000&0.8643\\
  0.5391&0.8643&1.0000
  \end{bmatrix}$ \tabularnewline \hline
\hline PLANCK+WP      & $301.60\pm 0.18$  &   $1.7442\pm 0.0093$  & $1089.86\pm 0.58$  &
  $\begin{bmatrix}
  1.0000&0.5698&0.5248\\
  0.5698&1.0000&0.8889\\
  0.5248&0.8889&1.0000
  \end{bmatrix}$ \tabularnewline \hline
\hline $\tabincell{c}{PLANCK+WP\\+BKP}$      & $301.55\pm 0.19$  &   $1.7407\pm 0.0097$  & $1089.68\pm 0.59$  &
  $\begin{bmatrix}
  1.0000&0.5519&0.5014\\
  0.5519&1.0000&0.8899\\
  0.5014&0.8899&1.0000
  \end{bmatrix}$ \tabularnewline \hline
\hline WMAP9+BKP      & $301.97\pm 0.65$  &   $1.7251\pm 0.0173$  & $1088.89\pm 0.91$  &
  $\begin{bmatrix}
  1.0000&0.3789&0.5014\\
  0.3789&1.0000&0.8611\\
  0.5014&0.8611&1.0000
  \end{bmatrix}$ \tabularnewline \hline
\end{tabular}
\caption{\label{tab5}The mean values, standard deviations of $\{l_A, R, z_*\}$ and the
         correlation matrix for the $\Omega_k$+$w$CDM model.}
\end{table}

In the CMB measurement, the distance to the last scattering surface can be accurately determined
from the locations of peaks and troughs of acoustic oscillations.
There are two quantities: one is the ``acoustic scale"
\begin{equation}
l_A=(1+z_*)\frac{\pi D_A(z_*)}{r_s(z_*)},
\end{equation}
and the other is the ``shift parameter"
 \begin{equation}
R=\sqrt{\Omega_{m0} H_0^2}(1+z_*)D_A(z_*).
\end{equation}
Here $D_A(z)= r(z)/(1+z)$ is the angular diameter distance and $z_*$ is the redshift at the last scattering
surface~\cite{Hu:1995en}
\begin{equation}
z_*=1048[1+0.00124(\Omega_{b0} h^2)^{-0.738}][1+g_1(\Omega_{m0} h^2)^{g_2}],
\end{equation}
where
\begin{equation}
\begin{split}
&g_1=\frac{0.0783(\Omega_{b0} h^2)^{-0.238}}{1+39.5(\Omega_{b0} h^2)^{0.763}},\\
&g_2=\frac{0.560}{1+21.1(\Omega_{b0} h^2)^{1.81}}.
\end{split}
\end{equation}
These quantities can be used to constrain some cosmological parameters
without need to use the full likelihoods of WMAP9~\cite{Hinshaw:2012aka} or Planck data~\cite{Ade:2013zuv}.

Based on the original idea proposed in~\cite{Wang:2007mza},
Hinshaw {\it et al.}~\cite{Hinshaw:2012aka} obtained constraints on the parameter combination ${\{l_A, R, z_*\} }$
from WMAP9 data based on the $w$CDM model without assuming a flat universe.
Wang and Wang~\cite{Wang:2013mha} obtained the mean values and normalized covariance matrix of
${\{l_A, R, \Omega_{b0} h^2,n_s\}}$ from WMAP9 and Planck+WP+Lensing data, respectively,
based on the $\Omega_k$+$\Lambda$CDM model.
Recently, Shafer and Huterer~\cite{Shafer:2013pxa} derived the related results about ${\{l_A, R, z_*\} }$ from WMAP9 and Planck+WP data, respectively, based on the flat $w$CDM model.
In this section, following Wang and Wang~\cite{Wang:2013mha} we obtain the Markov chains using the Markov Chain Monte Carlo sampler as implemented in the
CosmoMC package~\cite{Lewis:2002ah} and then derive constraints on the parameter combination ${\{l_A, R, z_*\}}$.
In our analysis, we focus on four cosmological models listed in Table~\ref{tab1}, based on the six-parameter model in the case of the flat $\Lambda$CDM model,  described by
\begin{equation}
\{\Omega_{b0}h^2,\Omega_{ dm0}h^2,\Theta_s,\tau,A_s,n_s\}, \nonumber
\end{equation}
where $\Theta_s$ is the ratio of the sound horizon to the angular diameter distance at the
photon decoupling, $\tau$ is the Thomson scattering optical depth due to reionization,
$A_s$ is the amplitude of primordial curvature perturbations and $n_s$ is the scalar spectral index.
The CMB data sets used in our analysis are listed in Table~\ref{tab1}.
Here we emphasize that both the tensor-to-scalar ratio $r$ and running of the scalar spectral index $\alpha_s$ are
allowed to vary if the BICEP2 $B$-mode polarization data are included,
because the $B$-mode power spectrum from the BICEP2 experiment implies the detect of primordial gravitational wave
at 7.0 $\sigma$ ignoring foreground dust~\cite{Ade:2014xna}
and allowing the running of the scalar spectral index reconciles the tension with the Planck constraints on $r$~\cite{hub1404}.
These results have been confirmed by other data on the same field from the
successor experiment Keck Array~\cite{Ade:2015fwj}.
However, it is argued  in~\cite{Mortonson:2014bja,Flauger:2014qra} that given
the uncertainties of the amplitude of the dust polarization at the BICEP2 frequency of 150 GHz one cannot say conclusively
at present whether the B-modes detected by BICEP2 are due to gravitational waves or just polarized dust.
 By using genus statistics, the authors of Ref~\cite{Colley:2014nna} claim to find the evidence for the primodal gravitational wave signal with $r = 0.11\pm0.04$.
 Planck team~\cite{Adam:2014bub} released the polarization data from 100 to 353 GHz,
extrapolation of the Planck 353 GHz data to 150 GHz gives a dust power, which is the same magnitude as reported by BICEP2.
Recently, Ref.~\cite{Ade:2015tva} performed a joint analysis of BICEP2/Keck and Planck data and obtained an upper
limit $r <0.12$ at 95\% confidence, showing little evidence of detection of primordial gravitational wave.
Therefore the tensor-to-scalar ratio is allowed to vary if the BKP data are included in our analysis.


The mean values, the standard deviations and their correlation matrix of ${\{l_A, R, z_*\}}$ (or ${\{l_A, R, \Omega_{b0} h^2\}}$)
for four cosmological models by using different data
are summarized in Table~\ref{tab2} to Table~\ref{tab5}, respectively.
It is noticed that Wang and Wang~\cite{Wang:2013mha} used $\Omega_{b0} h^2$ instead of $z_*$,
which gives identical constraints by replacing $\Omega_{b0} h^2$ with $z_*$.
Different from other cases in the third row of Table~\ref{tab1}, they also used Planck data together with Planck lensing.

From Table~\ref{tab2} to Table~\ref{tab5}, we see that the Planck data give tighter constraints on $\{l_A, R, z_*\}$
than WMAP9 in the same cosmological model. Including the BKP data does not change the results significantly but
the standard deviations seem to be a little larger, this is
because tensor perturbations have been considered
when we use Planck+WP+BKP to obtain constraints on $\{l_A, R, z_*\}$.
We also notice that there is some tension between the WMAP9 data and Planck data constraining on $\{l_A, R, z_*\}$.
For example, in Table~\ref{tab2} the constraints on $R$ ($z_*$) are inconsistent at about 1.5 $\sigma$ (1.6 $\sigma$) when
using WMAP9 and Planck+WP. The estimates of $\{l_A, R, z_*\}$ are consistent with each other within 1 $\sigma$ for the
$\Lambda$CDM model when we use Planck+WP and Planck+WP+BKP data.

Moreover, the difference of values of the reduced CMB data derived from same data for different models is not significant.
As stated in Ref.~\cite{Corasaniti:2007rf}, curvature or slightly modifying the dark energy parameters does
not significantly change the values of $\{l_A, R\}$. For example, from Table~\ref{tab2} and Table~\ref{tab3} we find that
the WMAP9 data give values of $\{l_A, R\}=\{301.95\pm 0.66,~1.7257\pm 0.0165\}$ ($\{302.02\pm 0.66,~1.7327\pm 0.0164\}$) for
the $\Lambda$CDM model without (with) spatial curvature which shows no significant difference.

\section{Cosmological parameters \label{sec:results}}

\begin{center}
 \begin{figure}
 \includegraphics[width=5in]{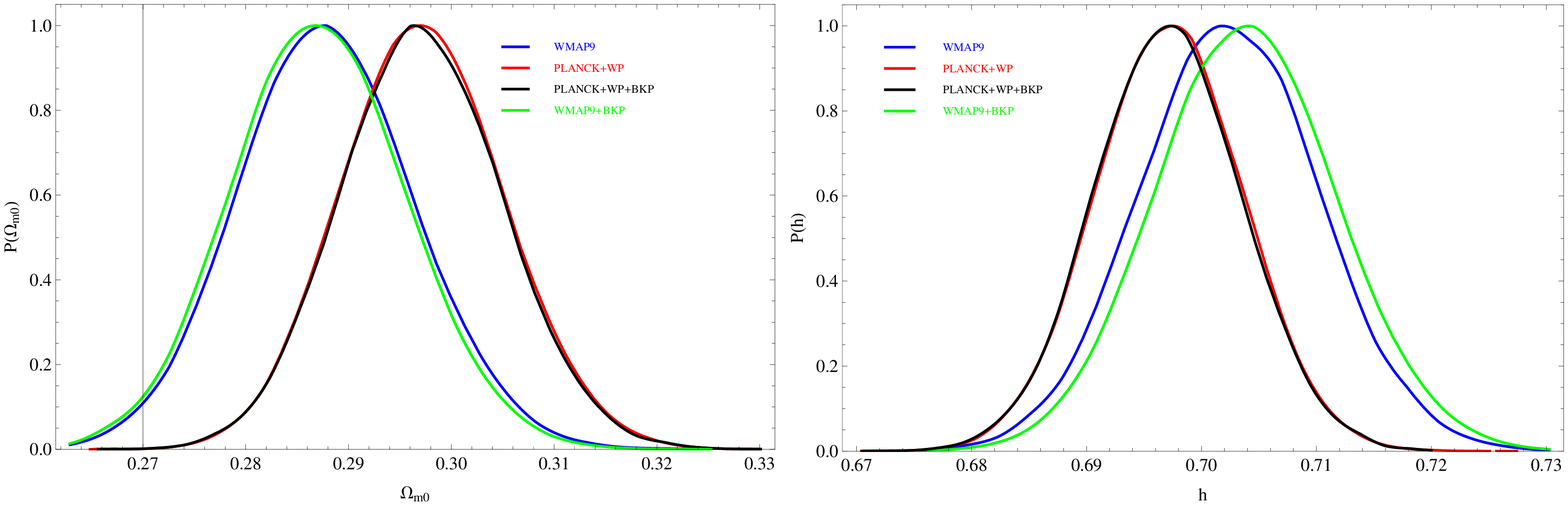}
 \caption{\label{fig:1}Marginalized posterior distributions for $h$ (right) and
  $\Omega_{m0}$ (left) of the flat $\Lambda$CDM model. }
\end{figure}
\end{center}

\begin{center}
 \begin{figure}
 \includegraphics[width=5in]{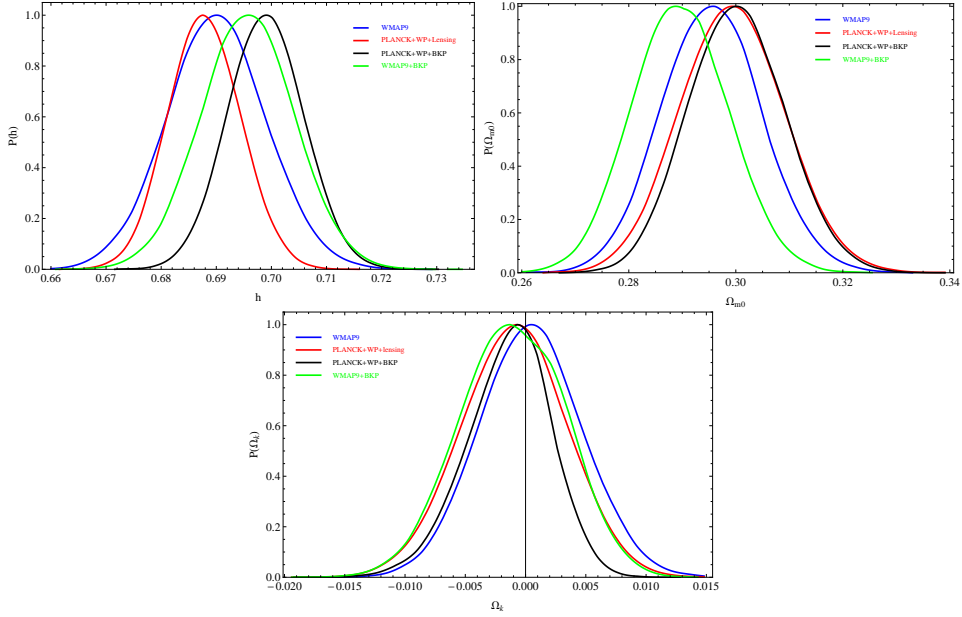}
 \caption{\label{fig:2}Marginalized posterior distributions for $h$ (top left), $\Omega_{m0}$ (top right) and $\Omega_k$ (bottom) of the $\Omega_k$+$\Lambda$CDM model. }
\end{figure}
\end{center}

\begin{center}
 \begin{figure}
 \includegraphics[width=5in]{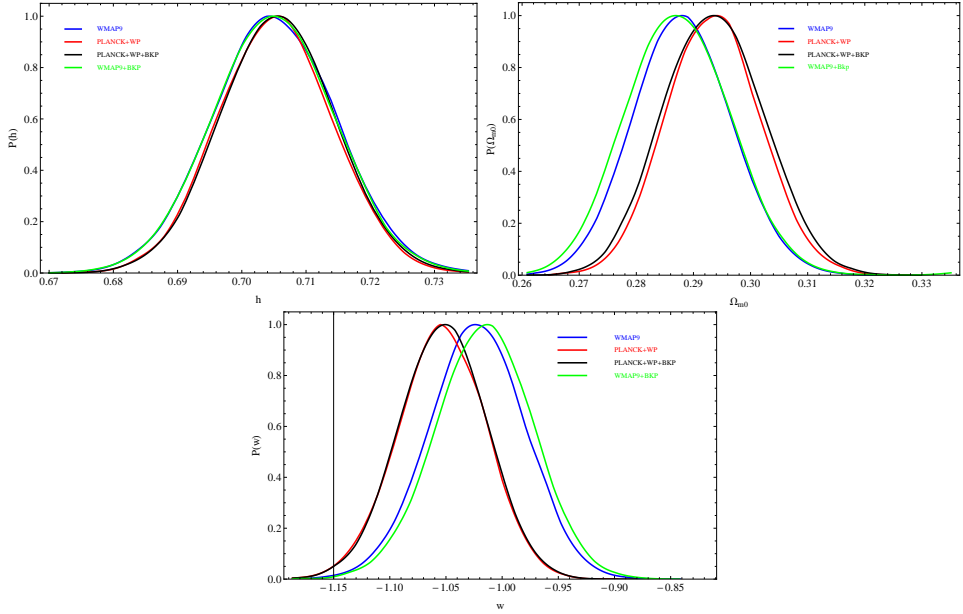}
 \caption{\label{fig:3}Marginalized posterior distributions for $h$ (top left), $\Omega_{m0}$ (top right) and $w$ (bottom) of the flat $w$CDM model.  }
\end{figure}
\end{center}

\begin{center}
 \begin{figure}
 \includegraphics[width=5in]{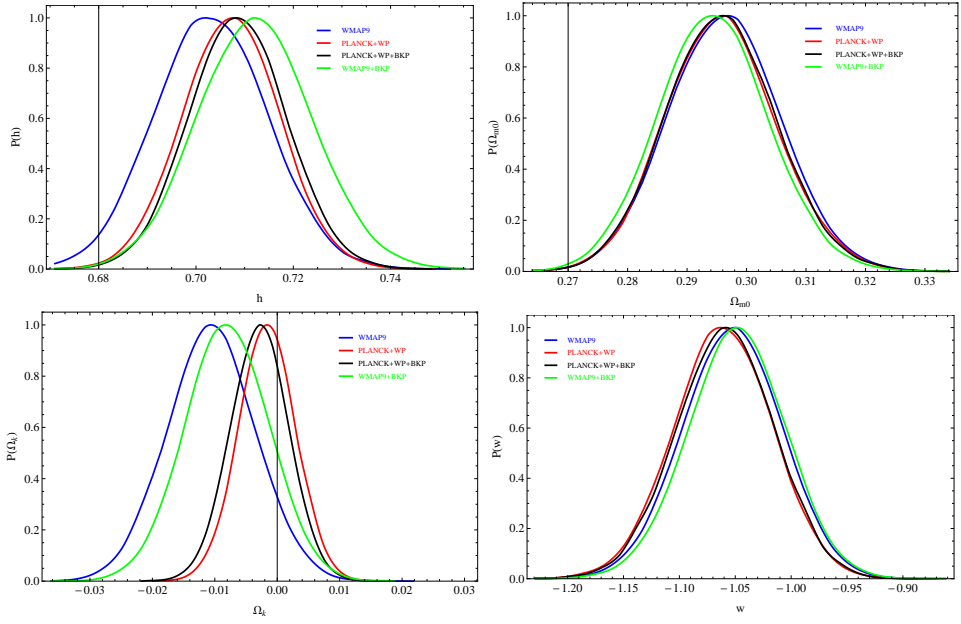}
 \caption{\label{fig:4}Marginalized posterior distributions for $h$ (top left), $\Omega_{m0}$ (top right), $\Omega_k$ (bottom left) and $w$ (bottom right) of the $\Omega_k$+$w$CDM model.  }
\end{figure}
\end{center}

In  the previous section  we have derived the reduced CMB data from WMAP9, Planck+WP, Planck+WP+BKP and WMAP9+BKP for
the $\Lambda$CDM model and the $w$CDM model with and without spatial curvature, respectively. In this section,
we focus on constraints on the cosmological parameters for the corresponding cosmological
models from reduced CMB data in combination with the low-redshift observational
data including the Union2.1 SNe Ia sample, Hubble parameter and BAO data, which are described in
 Appendix.
The best-fitted
values of $\Omega_{m0}$ and $h$ for the $\Lambda$CDM model, $\Omega_{m0}$, $h$ and $w$ for the
$w$CDM model and their 68\%  confidence level (CL) errors are given by using the
Markov Chain Monte Carlo analysis in the multidimensional parameter space in a Bayesian framework.
The results are summarized in
Table~\ref{tab6} to Table~\ref{tab9}, and their likelihoods are shown in Figure~\ref{fig:1} to Figure~\ref{fig:4}, respectively.

From Table~\ref{tab6} we  see that in the context of the flat $\Lambda$CDM model, the combination of Planck data favors
a relatively higher value of $\Omega_{m0}$ and a lower value of $h$ compared to the combination of WMAP9 data.
However, the reduced CMB data from Planck+WP do not lead to significantly improve the constraint on dark energy together with low-redshift observational data, compared
to the reduced CMB data from WMAP9 even though the Planck measures all of the CMB distance parameters $\{l_A, R, z_*\}$ more precisely, whose
errors are $2-3$ times smaller. This is because the Planck data appear to favor a higher  value of $\Omega_{m0}$ and a lower value of
$H_0$ in the standard six-parameter $\Lambda$CDM model, which are in tension with the magnitude-redshift relation for SNe Ia and recent
direct measurements of $H_0$~\cite{Ade:2013zuv}~\cite{Planck:2015xua}.
The constraints with BKP data suppress the value of $\Omega_{m0}$ and raise the value of $h$. The tendence seems to appear in
all the cosmological models we are considering here. These estimates of $\Omega_{m0}$ and $h$ are consistent with each other
within 1 $\sigma$ CL, but are in tension with the results derived by Planck~\cite{Ade:2013zuv}.

As we can see from Table~\ref{tab7}, in the context of $\Lambda$CDM model with spatial curvature, the constraints with Planck+WP+Lensing
give $h=0.6880^{+0.0090}_{-0.0096}$, which is inconsistent with the value $h=0.6988^{+0.0090}_{-0.0096}$ derived from Planck+WP+BKP
at about 1.2 $\sigma$ CL. The Planck team gives $h=0.6781$ from TT+LowP+Lensing and $h=0.6731$ from TT+LowP, and
the value of $h$ may be enhanced by Planck lensing data~\cite{Planck:2015xua}, so we can conclude that the constraints with Planck+WP in the
$\Lambda$CDM model with spatial curvature may give a lower value of $h$, which deviates more from $h=0.6988$.
Furthermore, the WMAP9 data favor a positive $\Omega_k$ (an open universe), but the Planck data give a negative $\Omega_k$ (a closed universe). However, there is no evidence for any departure from a
spatially flat geometry in these three cases.

A cosmological constant has an equation of state (EOS) $w=-1$. If we release the EOS $w$ of dark energy, the constraints with Planck+WP give
$w=-1.0507^{+0.0469}_{-0.0507}$, as shown in Table~\ref{tab8}, which favors the phantom region at 1 $\sigma$ CL. The combination with
BKP data gives a relatively higher value of $w$ ($w=-1.0489^{+0.0552}_{-0.0501}$), while the constraints with WMAP9 (WMAP9+BKP)
give $w=-1.0180^{+0.0535}_{-0.0667}$ ($w=-1.0155^{+0.0584}_{-0.0586}$).
All the three cases are consistent with the $\Lambda$CDM model.

WMAP9, Planck+WP, Planck+WP+BKP and WMAP9+BKP all give negative $\Omega_k$ (a closed universe) for the $\Omega_k$+$w$CDM model,
as shown in Table~\ref{tab9}, but are consistent with a flat geometry within 1 $\sigma$ CL.

\begin{center}
 \begin{figure}
 \includegraphics[width=5in]{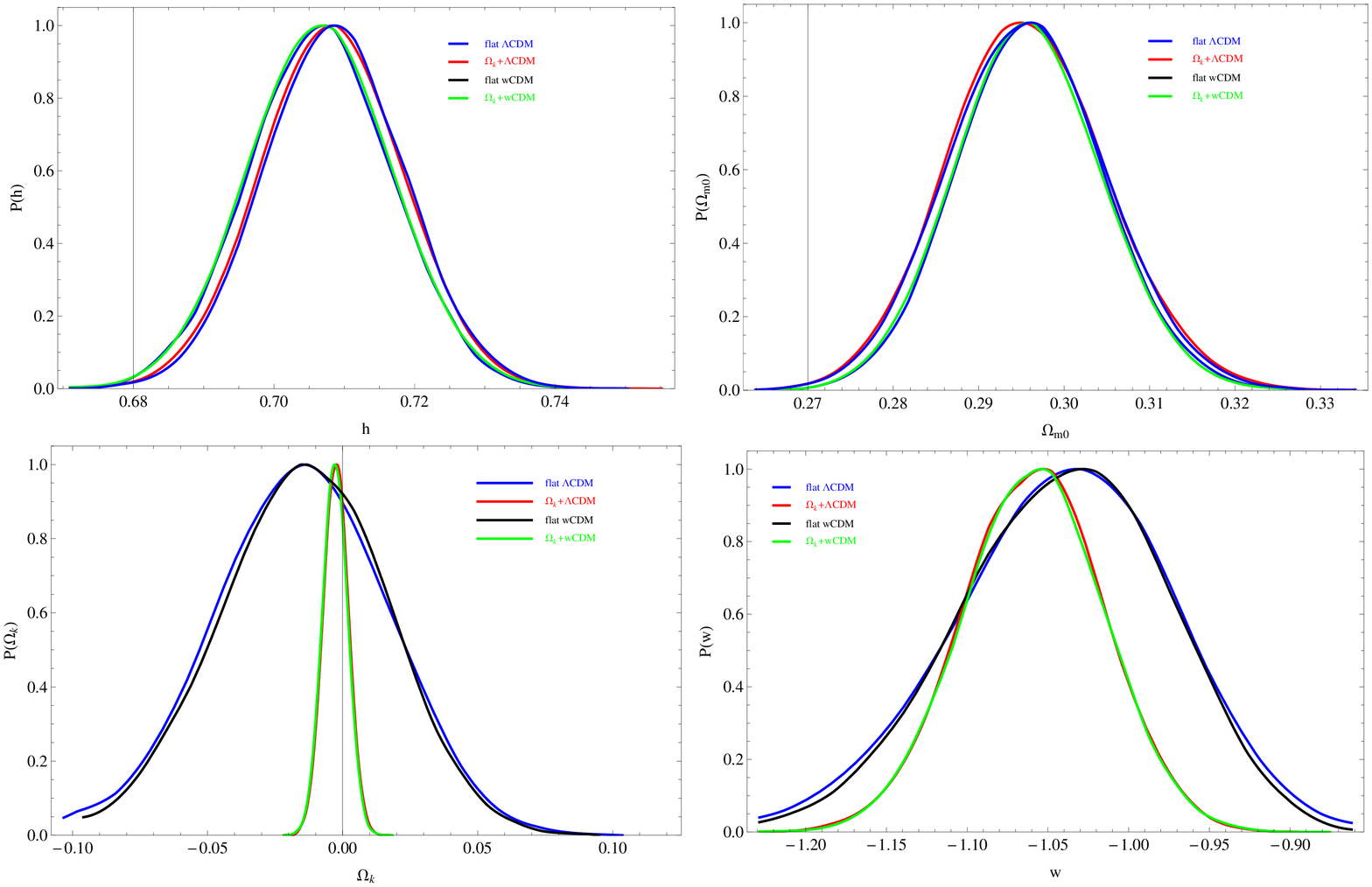}
 \caption{\label{fig:5}Marginalized posterior distributions for $h$ (top left), $\Omega_{m0}$ (top right), $\Omega_k$ (bottom left) and $w$ (bottom right) of the $\Omega_k$+$w$CDM model
 from Planck+WP+BKP. Different colors correspond to the different models used to derive the reduced CMB data from Planck+WP+BKP. }
\end{figure}
\end{center}

We also give the constraints for the $\Omega_k+ w$CDM model by using four sets of the reduced CMB data from Planck+WP+BKP derived from flat $\Lambda$CDM, $\Omega_k+\Lambda$CDM, flat $w$CDM and $\Omega_k+ w$CDM, respectively, together with the low-redshift observational data. The purpose is to see whether
the constraints on the cosmological parameters are sensitive to the choice of the reduced CMB data derived from different cosmological models. Their likelihoods are shown in Figure~\ref{fig:5}.
We can see that the likelihoods for the parameters $h$ and $\Omega_{m0}$ are almost the same in the four cases.
 As for the parameters $\Omega_k$ and $w$, the likelihoods show no significant difference between the flat $\Lambda$CDM and flat $w$CDM cases,  and between the $\Omega_k+\Lambda$CDM and $\Omega_k+ w$CDM cases, respectively. However, it can be seen clearly that
 the reduced CMB data derived from the cosmological model with spacial curvature can give much better
constraints on $\Omega_k$ than those derived from flat cosmological model.  But the values of
$\Omega_k$ and $w$ are still consistent with each other within 1 $\sigma$ CL in the four cases.


\begin{table}[!th]
\begin{tabular}{l|l|l}
\hline Data  & $\Omega_{m0}$  & $h$  \tabularnewline \hline
\hline WMAP9                       & $0.2877^{+0.0111}_{-0.0110}$ & $0.7022^{+0.0110}_{-0.0095}$ \tabularnewline \hline
\hline Planck+WP                   & $0.2967^{+0.0106}_{-0.0099}$ & $0.6975^{+0.0076}_{-0.0087}$ \tabularnewline \hline
\hline Planck+WP+BKP            & $0.2967^{+0.0109}_{-0.0097}$ & $0.6973^{+0.0079}_{-0.0091}$ \tabularnewline \hline
\hline WMAP9+BKP                & $0.2866^{+0.0116}_{-0.0101}$ & $0.7039^{+0.0097}_{-0.0101}$ \tabularnewline \hline
\end{tabular}
\caption{\label{tab6}Constraints with 1 $\sigma$ errors on
$\Omega_{m0}$ and $h$ for the flat $\Lambda$CDM model from SNe
Ia, HUB, BAO and reduced CMB data.}
\end{table}

\begin{table}[!th]
\begin{tabular}{l|l|l|l}
\hline Data  & $\Omega_{m0}$  & $h$  & $\Omega_k$\tabularnewline \hline
\hline WMAP9                       & $0.2946^{+0.0114}_{-0.0112}$ & $0.6904^{+0.0112}_{-0.0126}$ & $0.0007^{+0.0051}_{-0.0063}$ \tabularnewline \hline
\hline Planck+WP+Lensing           & $0.2986^{+0.0129}_{-0.0134}$ & $0.6880^{+0.0090}_{-0.0096}$ & $-0.0006^{+0.0054}_{-0.0067}$ \tabularnewline \hline
\hline Planck+WP+BKP            & $0.2963^{+0.0147}_{-0.0086}$ & $0.6988^{+0.0102}_{-0.0085}$ & $-0.0001^{+0.0038}_{-0.0055}$ \tabularnewline \hline
\hline WMAP9+BKP                & $0.2904^{+0.0100}_{-0.0143}$ & $0.6990^{+0.0079}_{-0.0146}$ & $-0.0013^{+0.0066}_{-0.0057}$ \tabularnewline \hline
\end{tabular}
\caption{\label{tab7}Constraints with 1 $\sigma$ errors on
$\Omega_{m0}$, $h$ and $\Omega_k$ for the $\Omega_k$+$\Lambda$CDM model from SNe
Ia , HUB, BAO and reduced CMB data.}
\end{table}

\begin{table}[!th]
\begin{tabular}{l|l|l|l}
\hline Data  & $\Omega_{m0}$  & $h$  & $w$\tabularnewline \hline
\hline WMAP9                       & $0.2878^{+0.0116}_{-0.0101}$ & $0.7043^{+0.0134}_{-0.0116}$ & $-1.0180^{+0.0535}_{-0.0667}$ \tabularnewline \hline
\hline Planck+WP                   & $0.2936^{+0.0103}_{-0.0110}$ & $0.7048^{+0.0123}_{-0.0104}$ & $-1.0507^{+0.0469}_{-0.0507}$ \tabularnewline \hline
\hline Planck+WP+BKP               & $0.2930^{+0.0115}_{-0.0106}$ & $0.7054^{+0.0122}_{-0.0118}$ & $-1.0489^{+0.0552}_{-0.0501}$ \tabularnewline \hline
\hline WMAP9+BKP                   & $0.2866^{+0.0113}_{-0.0100}$ & $0.7049^{+0.0126}_{-0.0119}$ & $-1.0155^{+0.0584}_{-0.0586}$ \tabularnewline \hline
\end{tabular}
\caption{\label{tab8}Constraints with 1 $\sigma$ errors on
$\Omega_{m0}$, $h$ and $w$ for the flat $w$CDM model from SNe
Ia , HUB, BAO and reduced CMB data.}
\end{table}

\begin{table}[!th]
\begin{tabular}{l|l|l|l|l}
\hline Data  & $\Omega_{m0}$  & $h$  & $w$  & $\Omega_k$\tabularnewline \hline
\hline WMAP9                       & $0.2963^{+0.0163}_{-0.0160}$ & $0.7007^{+0.0232}_{-0.0192}$ & $-1.0536^{+0.0791}_{-0.0720}$ & $-0.0091^{+0.0104}_{-0.0129}$ \tabularnewline \hline
\hline Planck+WP                   & $0.2945^{+0.0162}_{-0.0160}$ & $0.7078^{+0.0153}_{-0.0174}$ & $-1.0614^{+0.0833}_{-0.0740}$ & $-0.0013^{+0.0070}_{-0.0090}$ \tabularnewline \hline
\hline Planck+WP+BKP               & $0.2940^{+0.0164}_{-0.0122}$ & $0.7098^{+0.0137}_{-0.0175}$ & $-1.0576^{+0.0686}_{-0.0726}$ & $-0.0029^{+0.0081}_{-0.0078}$ \tabularnewline \hline
\hline WMAP9+BKP                   & $0.2936^{+0.0171}_{-0.0171}$ & $0.7105^{+0.0203}_{-0.0160}$ & $-1.0437^{+0.0661}_{-0.0776}$ & $-0.0066^{+0.0100}_{-0.0133}$ \tabularnewline \hline
\end{tabular}
\caption{\label{tab9}Constraints with 1 $\sigma$ errors on
$\Omega_{m0}$, $h$, $w$ and $\Omega_k$ for the $\Omega_k$+$w$CDM model from SNe
Ia, HUB, BAO and reduced CMB data.}
\end{table}

\section{Conclusions \label{sec:conclusions}}

The reduced CMB data provide an efficient summary of CMB information, and can be used to constrain cosmological parameters instead of the
full CMB power spectra. We have obtained the reduced CMB data from WMAP9 data
and Planck date based on the $\Lambda$CDM model and $w$CDM model with a flat
or curved space curvature, respectively. We have also used the BKP data
together with the WMAP9 and Planck data to derive the reduced CMB data. We have found that the Planck data give
tighter constraints on $\{l_A, R, z_*\}$ than WMAP9 in the same cosmological model.
While including the BKP data, the standard deviations seem to be a little larger
because of additional free parameters, the tensor-to-scalar ratio and running of the scalar spectral index.

We have combined these reduced CMB data with low-redshift observational data to constrain
the cosmological parameters for the $\Lambda$CDM model and the $w$CDM model.  The reduced
CMB data from Planck+WP do not lead to significant improvement to the constraint on dark energy  together with
low-redshift observational data, compared to the reduced CMB data from WMAP9.
Including BKP data results in a higher value of the Hubble constant especially
when the equation of state of dark energy and curvature are allowed to vary.
For the $\Omega_k$+$\Lambda$CDM model, the constraint from Planck+WP+Lensing in combination with low-redshift observations
gives $h=0.6880^{+0.0090}_{-0.0096}$, which is inconsistent with the value of $h=0.6988^{+0.0090}_{-0.0096}$
derived from Planck+WP+BKP at about 1.2 $\sigma$ CL.
The constraint on $w$ with Planck+WP gives $w=-1.0507^{+0.0469}_{-0.0507}$, favoring the phantom region at 1 $\sigma$ CL,
for a flat $wCDM$ model. The constraint on $w$ with WMAP9 or Planck+WP+BKP is consistent
with $w=-1$.
We have also found that the constraints on the cosmological parameter $\Omega_k$ and $w$ are a little sensitive to the choice of the reduced CMB data derived from different cosmological models.


\begin{acknowledgments}
This work was supported in part by the National Natural Science Foundation of China
(No.10821504, No.10975168, No.11035008, No.11175225 and No.11335012), and in part by the Ministry of Science and
Technology of China under Grant No. 2010CB833004 and No. 2010CB832805. RGC is
also supported by the Strategic Priority Research Program
¡°The Emergence of Cosmological Structures¡± of the
Chinese Academy of Sciences, Grant No. XDB09000000.
We used CosmoMC and CAMB.
We acknowledge the use of the WMAP, Planck and BICEP2 data and the Lenovo DeepComp 7000 supercomputer in SCCAS.
\end{acknowledgments}


\begin{appendices}

\subsection{Type Ia Supernovae}
In this work, we take the Union$2.1$
compilation~\cite{Union2.1}, which contains $580$ SNe Ia data over the
redshift range $0.015 \le z \le 1.414$. The chisquare is defined  as
\begin{equation}
\chi^2_{SN}=\sum_{i=1}^{580}\frac{[\mu^{obs}(z_i)-\mu^{th}(z_i)]^2}{\sigma_{SN}^2(z_i)} ,
\end{equation}
where $\mu^{obs}(z)$ is the measured distance modulus from the data and $\mu^{th}(z)$ is the theoretical distance modulus, defined as
\begin{equation}
\mu^{th}(z)=5\log_{10}{d_L}+\mu_0,~~\mu_0=42.384-5\log_{10}{h}.
\end{equation}
The luminosity distance is
\begin{equation}
d_L(z)=(1+z)r(z),
\end{equation}
where $r(z)$ is the comoving distance defined in equation~(\ref{eq:3}).   The nuisance
parameter $\mu_0$ can be eliminated by expanding $\chi^2$ with respect to
 $\mu_0$ as~\cite{Nesseris:2005ur} :
\begin{equation}
\chi^2_{SN}=A+2B\mu_0+C\mu^2_0 ,
\end{equation}
where
\begin{equation}
\begin{split}
&A=\sum_{i=1}^{N}\frac{[\mu^{th}(z_i;\mu_0=0)-\mu^{obs}(z_i)]^2}{\sigma_{SN}^2(z_i)},\\
&B=\sum_{i=1}^{N}\frac{\mu^{th}(z_i;\mu_0=0)-\mu^{obs}(z_i)}{\sigma_{SN}^2(z_i)},\\
&C=\sum_{i=1}^{N}\frac{1}{\sigma_{SN}^2(z_i)} .
\end{split}
\end{equation}
The \textbf{$\chi^{2}_{SN}$} has a minimum as
\begin{equation}
\tilde{\chi}^{2}_{SN}=A-B^2/C~,\label{sn}
 \end{equation}
In this way the nuisance parameter $\mu_{0}$ is removed. This technique is equivalent to performing a
uniform marginalization over $\mu_{0}$~\cite{Nesseris:2005ur}. We will adopt
\textbf{$\tilde{\chi}^{2}_{SN}$} as the goodness of fitting instead of \textbf{$\chi^{2}_{SN}$}.


\subsection{Observational Hubble parameter (HUB)}

In this paper we use 19 observational Hubble data over the redshift range: $0.07\leq z \leq 2.3$,
which contain 11 observational Hubble data obtained from the differential ages of
passively evolving galaxies~\cite{Simon:2004tf,Stern:2009ep}, and 8 $H(z)$ data at eight different redshifts
obtained from the differential spectroscropic evolution of early type galaxies as a function of redshift~\cite{Moresco:2012jh}.
The chisqure is defined as
\begin{equation}
\chi^{2}_{HUB}=\sum_{i=1}^{N}\frac{[H_{th}(z_i)-H_{obs}(z_i)]^2}{\sigma_H^2(z_i)},
\end{equation}
where $H_{th}(z)$ and $H_{obs}(z)$ are the theoretical and observed values of Hubble parameter,
and $\sigma_H$ denotes the error of observed data.


\subsection{Baryon Acoustic Oscillation (BAO)}

Baryon Acoustic Oscillation provides an efficient method for measuring the
expansion history of the universe by using features in the cluster
of galaxies with large scale survey. Here we use the results from
the following five BAO surveys: the 6dF Galaxy Survey, SDSS DR7,
SDSS DR9, WiggleZ measurements and the radial BAO measurement.

\subsubsection{6dF Galaxy Survey}

The 6dFGS BAO detection can constrain the distance-redshift
relation at $z_{eff}=0.106$~\cite{Beutler:2011hx}. 
it gives a measurement of the distance ratio
\begin{equation}
\frac{r_s(z_d)}{D_V(z=0.106)}=0.336\pm0.015,
\end{equation}
where $r_s(z_d)$ is the comoving sound horizon at the baryon drag epoch
when baryons became dynamically decoupled from photons.
The redshift $z_d$ is well approximated by~\cite{Eisenstein:1997ik}
\begin{equation}
z_d=\frac{1291(\Omega_{m0} h^2)^{0.251}}{1+0.659(\Omega_{m0} h^2)^{0.828}} [1+b_1(\Omega_{b0} h^2)^{b_2}],
\end{equation}
where
\begin{equation}
\begin{split}
&b_1=0.313(\Omega_{m0} h^2)^{-0.419}[1+0.607(\Omega_{m0} h^2)^{0.674}],\\
&b_2=0.238(\Omega_{m0} h^2)^{0.223}.
\end{split}
\end{equation}
The effective ``volume'' distance $D_V$ is a combination of the angular-diameter distance $D_A(z)$ and the Hubble parameter $H(z)$,
\begin{equation}
\begin{split}
D_V(z)&=\left [(r(z))^2\frac{z}{H(z)}\right ]^{1/3}\\
&=[(1+z)^2D_A(z)^2\frac{z}{H(z)}]^{1/3}.
\end{split}
\end{equation}
The $\chi^2_{6dF}$ is defined by
\begin{equation}
\chi^{2}_{6dF}=\frac{[({r_s(z_d)}/{D_V(0.106))_{th}}-0.336]^2}{0.015^2}.
\end{equation}

\subsubsection{SDSS DR7}

 The joint analysis of the 2-degree Field Galaxy Redshift
Survey data and the  Sloan Digital Sky Survey Data Release $7$ data
gives the distance ratio at $z=0.2$ and
$z=0.35$~\cite{Percival:2009xn}:
\begin{equation}
\begin{split}
\frac{r_s(z_d)}{D_V(z=0.2)}=0.1905\pm0.0061,\\
\frac{r_s(z_d)}{D_V(z=0.35)}=0.1097\pm0.0036.
\end{split}
\end{equation}
By applying the reconstruction technique~\cite{Eisenstein:2006nk} to the clustering of galaxies from
the SDSS DR7 Luminous Red Galaxies sample, and sharpening the BAO feature, Padmanabhan {\it et al.} obtained
 the distance ratio at $z=0.35$~\cite{Padmanabhan:2012hf} :
\begin{equation}
\frac{r_s(z_d)}{D_V(z=0.35)}=0.1126\pm0.0022.
\end{equation}
The SDSS DR7 and SDSS DR7 reanalysis results are based on the same survey
and the latter gives a higher precision than the former, we therefore take the SDSS DR7 reanalysis data instead of the first one.
The $\chi^2_{DR7-re}$ used in the Markov Chain  Monte Carlo analysis is
\begin{equation}
\chi^2_{DR7re}=\frac{[(\frac{r_s(z_d)}{D_V(0.35)})_{th}-0.1126]^2}{0.0022^2}.
\end{equation}

\subsubsection{SDSS DR9}

The SDSS DR9 measurement gives the distance ratio at $z=0.57$~\cite{Anderson:2012sa}:
\begin{equation}
\frac{r_s(z_d)}{D_V(z=0.57)}=0.0732\pm0.0012.
\end{equation}
The chisquare here is defined as
\begin{equation}
\chi^2_{DR9}=\frac{[(\frac{r_s(z_d)}{D_V(0.57)})_{th}-0.0732]^2}{0.0012^2}.
\end{equation}

\subsubsection{The WiggleZ measurements}

The WiggleZ team measures the acoustic parameter by encoding some shape information on the power
spectrum~\cite{Blake:2011en}:
\begin{equation}
A(z)=\frac{D_V(z)\sqrt{\Omega_{m0}H_0}}{z}.
\end{equation}
The baryon acoustic peaks measured at redshifts $z=0.44$,
 $0.6$ and $0.73$ in the galaxy correlation function of the final
dataset of the WiggleZ Dark Energy Survey give
\begin{equation}
\begin{split}
&A(z=0.44)=0.474\pm0.034,\\
&A(z=0.60)=0.442\pm0.020,\\
&A(z=0.73)=0.424\pm0.021.
\end{split}
\end{equation}
The corresponding chisquare is defined as
\begin{equation}
\chi^2_{Wig}=X^TV^{-1}X,
\end{equation}
where
\begin{equation}
X=
  \begin{bmatrix}
  A(z=0.44)_{th}-0.474\\
  A(z=0.60)_{th}-0.442\\
  A(z=0.73)_{th}-0.424
  \end{bmatrix},
\end{equation}
and its inverse covariance matrix is
\begin{equation}
V^{-1}=
  \begin{bmatrix}
   1040.3&-807.5&336.8\\
   -807.5&3720.3&-1551.9\\
   336.8&-1551.9&2914.9
  \end{bmatrix}.
\end{equation}

\subsubsection{Radial BAO}

The radial (line-of-sight) baryon acoustic scale can also be
measured by using the SDSS data. It is independent from the BAO
measurements described above.
 The measured quantity is
\begin{equation}
\bigtriangleup_z(z)=H(z)r_s(z_d),
\end{equation}
whose values are given by~\cite{Gaztanaga:2008de} as
\begin{equation}
\begin{split}
&\bigtriangleup_z(0.24)=0.0407\pm{0.0011},\\
&\bigtriangleup_z(0.43)=0.0442\pm{0.0015}.
\end{split}
\end{equation}

\subsection{Reduced CMB data}

The chisquare for the reduced CMB data is defined by
\begin{equation}
\chi^2_{CMB}=X^TC^{-1}X,
\end{equation}
where
\begin{equation}
X=
  \begin{bmatrix}
  (l_A)_{th}-(l_A)_{obs}\\
  R_{th}-R_{obs}\\
  (z_*)_{th}-(z_*)_{obs},
  \end{bmatrix}
\end{equation}
and $C$ is the related covariance matrix.

\end{appendices}


\end{document}